\documentclass[12pt]{iopart}

\usepackage{graphicx}
\begin{document}

\title[]{Double-diffusive convection and baroclinic instability in a differentially heated and initially stratified rotating system: the barostrat instability}

\author{Miklos Vincze$^{1,2}$\footnote{Corresponding 
author: vincze.m@lecso.elte.hu}, Ion Borcia$^1$, Uwe Harlander$^1$, \\and Patrice Le Gal$^3$}

\address{$^1$ Department of Aerodynamics and Fluid Mechanics, Brandenburg University of Technology (BTU) Cottbus-Senftenberg, Siemens-Halske-Ring 14, D-03046 Cottbus, Germany\\}
\address{$^2$ MTA-ELTE Theoretical Physics Research Group, P\'azm\'any P. s. 1/A, H-1117 Budapest, Hungary\\}
\address{$^3$ Institut de Recherche sur les Ph\'enom\`enes Hors Equilibre, CNRS - Aix-Marseille University - Ecole Centrale Marseille, 49 rue F. Joliot-Curie, 13384 Marseille, France}

\ead{vincze.m@lecso.elte.hu}

\begin{abstract}
A water-filled differentially heated rotating annulus with initially prepared stable vertical salinity profiles is studied in the laboratory. Based on two-dimensional horizontal particle image velocimetry (PIV) data, and infrared camera visualizations, we describe the appearance and the
characteristics of the baroclinic instability in this original configuration. First, we show that when the salinity profile is linear and confined between two non stratified layers at top and bottom, only two separate shallow fluid layers can be destabilized. These unstable layers appear nearby the top and the bottom of the tank with a stratified motionless zone between them.  This laboratory arrangement is thus
particularly interesting to model geophysical or astrophysical situations where stratified regions are often juxtaposed to convective ones. Then, for more general but stable initial density profiles, statistical measures are introduced to quantify the extent of the baroclinic instability at given depths and to analyze the connections between this depth-dependence and the
vertical salinity profiles. We find that, although the presence of stable stratification generally hinders full-depth overturning, double-diffusive convection can yield development of multicellular sideways convection in shallow layers and subsequently to a multilayered baroclinic instability. Therefore we conclude that by decreasing the characteristic vertical scale of the flow, stratification may even enhance the formation of cyclonic and anticyclonic eddies (and thus, mixing) in a local sense.
\end{abstract}

\maketitle

\section{Introduction}
The differentially heated rotating fluid annulus is a widely studied experimental apparatus
for modeling large-scale features of the mid-latitude atmosphere (planetary waves,
cyclogenesis, etc.). In the classic set-up, first studied by the group of 
Fultz \etal (1959) and, in parallel by Fowlis and Hide (1965), a rotating cylindrical tank is divided into three
coaxial sections: the innermost domain is kept at a lower temperature, whereas the
outer rim of the tank is heated. The working fluid in the annular gap in between thus
experiences lateral temperature difference $\Delta T$ in the radial direction, which initiates a `sideways-convective' circulation. These boundary conditions imitate the meridional temperature gradient of the terrestrial atmosphere between the poles and the equator. The Coriolis effect arising due to the rotation of the tank modifies this axisymmetric basic flow markedly, and leads first to the formation of an azimuthal flow and then to baroclinic instability with cyclonic and anticyclonic eddies in the full water depth. In the present work, we study a modified version of this
experiment, in which -- besides the aforementioned radial temperature difference -- vertical salinity stratification is also present.

Whereas the baroclinic instability experiments are classical models of mid-latitude atmospheric dynamics, our aim is to create a juxtaposition of convective and motionless stratified layers in the laboratory. 
We expect therefore that the lessons learned from our original experimental set-up may well be useful for future laboratory arrangements (in somewhat different geometries and material properties) that could mimic the convective and radiative zones of stars, the tropospheres and stratospheres of planetary atmospheres or the surface turbulent sea layers above deeper stratified waters. Let us stress in particular that the
exchange of momentum and energy between these layers (in particular by the propagation of internal gravity waves) is a major issue in astrophysics, in atmospheric sciences as well as in oceanography. Therefore, we claim that our present set-up may lead to
new insights in geo- and astrophysical applications (Plougonven and Snyder, 2007; Nettelmann \etal, 2015).

Small variations in the density of a fluid parcel ($\delta\rho$) are connected to those of salinity ($\delta S$) and temperature ($\delta T$) in the form of $\delta \rho=-\alpha \delta T + \beta \delta S$, where $\alpha$ and $\beta$ denote the coefficients of thermal expansion and saline contraction, respectively. The global thermohaline ocean circulation is driven by the endless competition between the thermal and saline contributions to the buoyancy of each and every parcel of seawater (Vallis, 2006). Better understanding the nature of this complex interplay has been the key motivation of the present study.

Large enough (stable) vertical salinity gradients may inhibit the formation of full-depth (unicellular) overturning flow, even in such laterally heated systems as the rotating annulus experiments.
The vertical extent $\lambda$ of a convective cell is naturally limited by the condition that the initial (saline) density difference between the top and bottom of the cell cannot exceed the (thermal) horizontal density difference between the lateral sidewalls, i.e.:
\begin{equation}
\langle\rho\rangle \alpha \Delta T \approx \lambda \Bigg|\frac{\partial\rho}{\partial z}\Bigg|,
\end{equation}
where $\langle \rho \rangle$ is the average density within the cell. This scale can be expressed with buoyancy frequency $N(z)=\sqrt{g \langle\rho(z)\rangle^{-1}\,|d\rho(z)/dz|}$, in the form
\begin{equation}
\lambda(z)=\frac{g\alpha\Delta T}{N^2(z)}.
\label{lambda}
\end{equation}
This characteristic vertical scale of (non-rotating) \emph{double-diffusive convection} is referred to as the Chen scale, since the pioneering experimental analysis of Chen \etal (1971). Within each cell the convective flow yields mixing and the initial stratification vanishes. Thus, a so-called `double-diffusive staircase' develops with jump-wise density changes between these locally homogenized cells. Double-diffusive staircases are formed in natural water bodies, ranging from saline lakes to oceans (Boehrer, 2012; Schmitt, 1994). However, due to the inevitable mixing between the cells, such configurations are mostly transient. Hopefully, as the spatial scales involved in the baroclinic as well as in the convective instabilities are quite large, the transient characteristic times calculated on these lengths are long enough to permit a full investigation. Some of the experimental runs that we will describe later lasted several days.

The paper is organized as follows: after having recalled the basic principles of double-diffusive convection (and its response to rotation) and then of the baroclinic instability,
in sections \ref{DDconv} and \ref{Baro} respectively, in section \ref{Setup} we describe our experimental arrangement and measurement methods. Section \ref{Results} is devoted to the presentation of the results and
their analysis that will concern two sets of experimental runs. In the first set (subsection \ref{simple}), the salinity profiles were mainly linear (with a single buoyancy frequency in the core of the fluid) with two non-stratified regions at the top and bottom of the full water layer. The corresponding results permit to test the possibility to confine the baroclinic instability in two shallow layers separated by a motionless zone. In a second part of this result section (subsection \ref{complex}), more complex salinity profile have been realized with vertical variations of the buoyancy frequency. In this case, multilayered convection sets in. The analysis is thus much more complex and necessitates a particular method to discriminate and quantify the dynamics inside these independent layers. Section \ref{Conclusion} summarizes our main results and opens discussions on the future possibilities of our original experimental set-up.

\section{Basics of double-diffusive dynamics and its response to rotation}
\label{DDconv}

\begin{figure}[!h]
\begin{center}
\noindent\includegraphics[width=10cm]{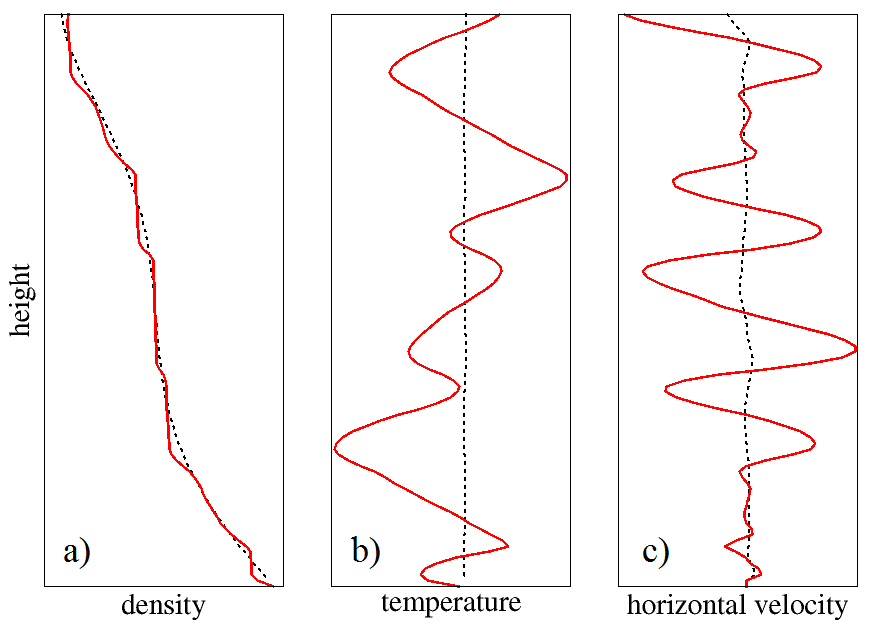}
\caption{Initial (black dotted) and final (red) vertical profiles of total density (a), temperature (b) and horizontal velocity (c) from a 2D, non-rotating simulation of double-diffusive convection, demonstrating the development of a double-diffusive `staircase'. Note the blocking at the regions of steeper density gradient. Units are arbitrary.}
\label{sim_profiles}
\end{center}
\end{figure}

For the sake of qualitative demonstration, in Fig. \ref{sim_profiles}a-c we present typical vertical
profiles of total density, temperature, and horizontal velocity $u$, from two-dimensional solutions of the non-rotating Boussinesq equations, acquired via {\it numerical simulation} using the open-source software package Advanced Ocean Modeling of 
K\"ampf (2010). No-flux conditions were prescribed at the top and bottom boundaries of the rectangular solution domain, and differential heating was applied at its lateral walls: the temperature at the vertical sidewall at the left (right) hand side of the domain was prescribed to be $\Delta T/2=4$ K colder (warmer) that the initially uniform temperature $T_0$ of the fluid. The fluid had a stable and continuous vertical density gradient and was at rest at time zero. Each profile of Fig. \ref{sim_profiles} was extracted from the vertical column at the middle of the domain. Since these graphs serve demonstration purposes only, no units are given.

In panel (a) it is clearly visible that the initial density stratification (dashed line) has developed into the aforementioned double-diffusive staircase (solid line, red online); see also the corresponding velocity profiles in panel c). In qualitative agreement with the formula (\ref{lambda}), the largest cells -- both in terms of vertical size $\lambda$ and of water flux -- appear at mid-depths, where the initial salinity profile exhibited the smallest gradient.
Note, however, that in the regions characterized by the largest density gradients, the flow is basically blocked (the horizontal velocities are practically zero): in the spirit of (\ref{lambda}), the large local $N$ would yield very shallow convective layers, but their formation is inhibited by viscous effects. This "viscous cut-off" bounds the cell thickness from below: it cannot be smaller then a certain $\lambda_{\rm crit}$, or, in terms of the buoyancy frequency, there is a maximum $N_{\rm crit}$, above which convection is locally blocked.

In their experiments, Chen \etal (1971) investigated the development of convective rolls in a rectangular vertical tank filled up with saline water of initially uniform vertical stratification with differential heating at two opposing lateral walls. His results are expressed in terms of a certain version of Rayleigh number $Ra_{\rm Chen}$, where the vertical scale is given by $\lambda$ -- as derived from (\ref{lambda}) -- and the characteristic temperature difference is measured between the lateral walls; thus $Ra_{\rm Chen}$ incorporates the effects of both the horizontal and the vertical density gradients:
\begin{equation}
Ra_{\rm Chen}\equiv \frac{g\alpha \Delta T}{\nu \kappa_T}\lambda^3=\frac{(g\alpha\Delta T)^4}{\nu \kappa_T N^6},
\label{Ra_Chen}
\end{equation}
where $\kappa_T$ is the thermal diffusivity and $\nu$ is the kinematic viscosity of the working fluid.

Chen \etal (1971) found that the critical value of $\lambda_{\rm crit}$ or, equivalently, $N_{\rm crit}$ is encountered around $Ra_{\rm Chen}^{\rm crit}\approx 15000$, below which convection is inhibited. If $Ra_{\rm Chen}$ exceeds this threshold, simultaneously or successively formed multicellular double-diffusive convection rolls can develop (Kranenborg and Dijkstra, 1995).

Introducing \emph{rotation} around the vertical axis modifies the dynamics of double-diffusive convection markedly, as demonstrated in the theoretical work of Kerr (1995).
His conceptual set-up involves vertical salinity gradient and differentially heated lateral sidewalls with gapwidth $d$, which rotate at angular velocity $\Omega$ (i.e. Coriolis parameter $f\equiv 2\Omega$). Unlike Chen, Kerr defines separate Rayleigh numbers for the vertical and horizontal directions ($H$ and $R$, respectively), both taken with the same length scale $d$, which is, in fact, the only geometrical scale here, since the vertical extent of the domain is assumed to be infinite in the calculations. The horizontal Rayleigh number $H$ reads as $H\equiv {g\alpha \Delta T d^3}/(\nu \kappa_T)$, whereas its vertical counterpart $R$ is given by
$R\equiv {N^2 d^4}/(\nu \kappa_T)$.
Note, that based on these formulae, the Chen scale $\lambda$ (defined in (\ref{lambda})), can be expressed as $\lambda = ({H}/{R})d$.

\begin{figure}[!h]
\begin{center}
\noindent\includegraphics[width=10cm]{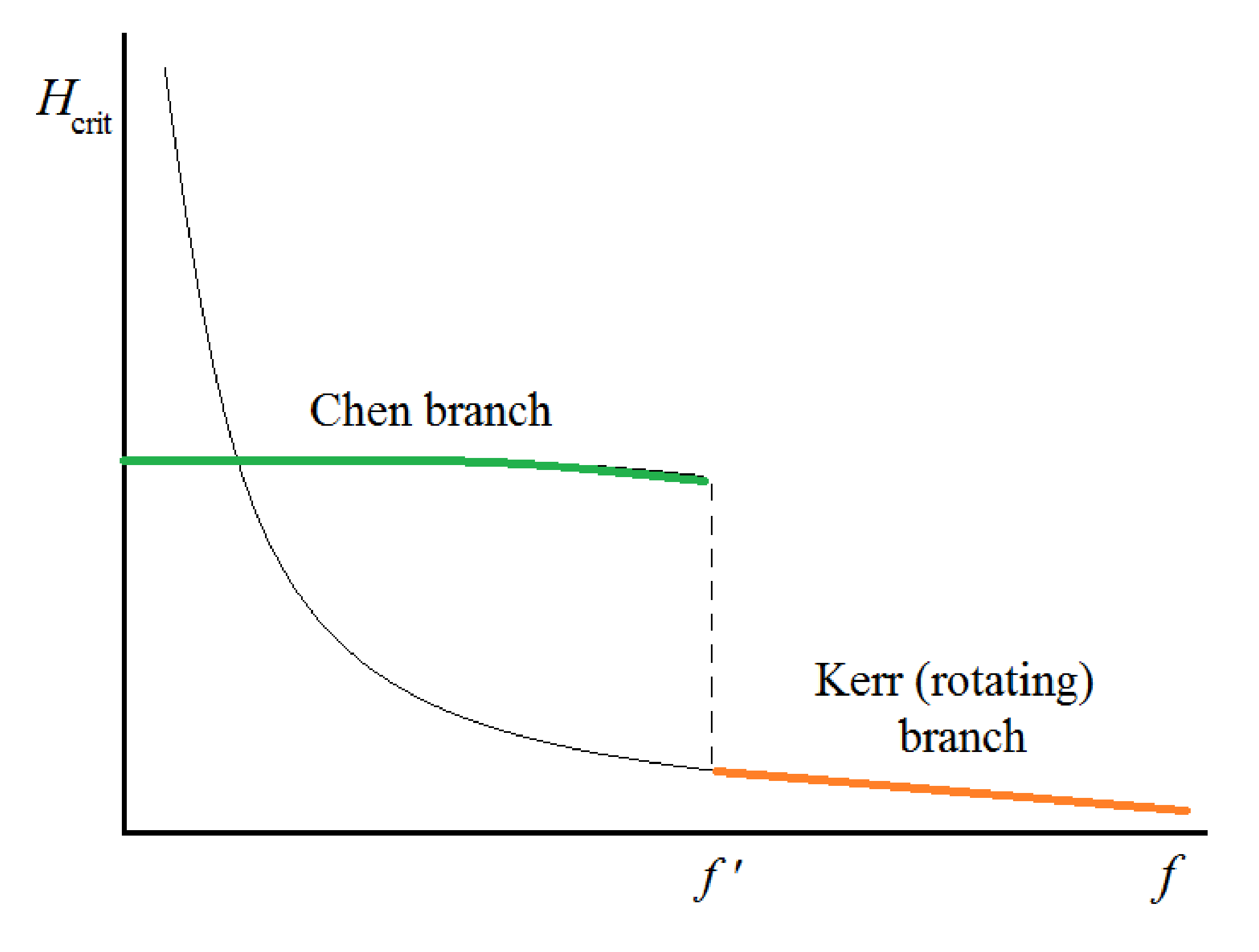}
\caption{A sketch of the stability diagram $H_{\rm crit}$ as a function of Coriolis parameter $f$, for a given $R$. For slow rotations, the slowly changing branch -- here, referred to as `Chen branch' -- is stable, but above a certain threshold $f'$  this branch vanishes, and only the sharply decreasing $H_{\rm crit} \sim f^{-1}$ Kerr's "rotating branch" survives.}
\label{bifurc}
\end{center}
\end{figure}

Kerr studied the $f$-dependence of the minimum $H_{\rm crit}$, below which the flow is blocked, at different fixed values of $R$ and found two types of solutions in all cases. A schematic sketch of a typical stability diagram is shown in Fig. \ref{bifurc}. One branch (green curve) is only slowly changing with $f$: this result is consistent with the findings of Chen \etal (1971) in the $f=0$ limit case.

This branch vanishes at a certain threshold $f'$, whose value depends on $R$: $f'(R)$ is an increasing function of $R$.
In the $f>f'$ regime the other branch of solutions describes the appearance of a new regime (orange curve in Fig. \ref{bifurc}). This branch decays in an inversely proportional manner with $f$ -- thus it is non-existent in the $f=0$ limit -- and its parameter $A$ depends on $R$ as well: $H_{\rm crit}(f) =A(R) \,f^{-1}$.

\begin{figure}[!h]
\begin{center}
\noindent\includegraphics[width=10cm]{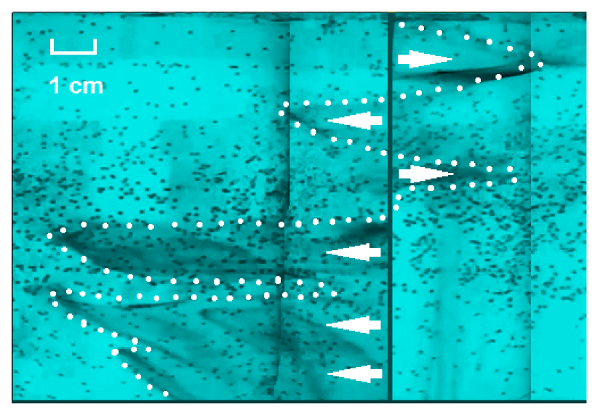}
\caption{Dye tracing in a rotating experiment. The stick from where the dye was released is the thick vertical silhouette. The propagating dye fronts are marked with white dots for better visibility.}
\label{dye_demo}
\end{center}
\end{figure}

For a qualitative understanding of the effect of rotation on the double-diffusive background flow let us consider an experiment with counter-clockwise, i.e. `Northern hemisphere'-like, rotation. As observed from the co-rotating frame of reference, the Coriolis force tends to push the horizontal flow branches to their right: at large enough values of rotation rate $\Omega$ and/or horizontal velocity $u$ this action can yield strong zonal flows in the system. Their directions are alternating with depth, corresponding to the convective branches of double-diffusive convective cells; the Coriolis effect yields \emph{zonal} propagation in the presence of \emph{radial} buoyancy gradient. This is demonstrated in the composite photo of Fig. \ref{dye_demo} taken during one of our experimental runs corresponding to the multi-layered cases that will be detailed in section \ref{complex}. The figure presents the side view (through the transparent glass walls of the rotating tank) of dye patterns propagating away from a painted wooden stick (black) in `eastward' (rightward) and `westward' (leftward) zonal directions at rotation rate $\Omega=2.5$ rpm, and average buoyancy frequency $\langle N \rangle =3.26$ rad/s.

It is important to emphasize, however, that the presence of baroclinic instability -- resulting in zonal wave propagation in the annular tank -- may largely add to the complexity of the emerging flow fields. Note also that even if Kerr (1995) never mentioned it, the solutions for $f>f'$ seem very close in shape and wavelength to the classical baroclinic eddies.

\section{Baroclinic instability}
\label{Baro}

The most important non-dimensional parameter, traditionally used to describe the flow regimes in a rotating system with a finite kinematic viscosity $\nu$ is the Taylor number $Ta$, which reads as follows:
\begin{equation}
Ta=\frac{\Omega^2}{h}\frac{4 d^5}{\nu^2},
\label{Ta}
\end{equation}
where $h$ is the vertical scale and $d$ is the `gapwidth' of the apparatus (or more generally, the horizontal scale), as in the earlier formulae.

Another relevant parameter, the thermal Rossby number $Ro_T$ is defined as the ratio of the characteristic velocity of the flow -- driven by the relative density change in the lateral direction, $\alpha\Delta T$ -- to the rotation rate, in the form of
\begin{equation}
Ro_T=\frac{h}{\Omega^2}\frac{g\alpha\Delta T}{d^2},
\label{Ro}
\end{equation}
where $g$ is the acceleration due to gravity. Comparing the above formulae it is obvious that for a fixed $\Delta T$ (as in the experiments to be discussed here), $Ta$ and $Ro_T$ are inversely proportional to each other and therefore the combination $\Omega^2/h$ can sufficiently parametrize the flow in the given experimental setup.

In a certain region of the $Ro_T$--$Ta$ parameter space, the slope of the isopycnal surfaces in the water (which are not horizontal in this rotating system, due to thermal wind balance) surpasses a critical threshold and {\it baroclinic instability} emerges (Vallis, 2006). The extra potential energy stored in these tilted layers is then released via the excitation of complex wavy patterns, consisting of meandering zonal jets, accompanied by cyclonic and anticyclonic eddies.
For $\Delta T= 6$ K (that is the typical case for this study, as will be discussed later) the Taylor number corresponding to the onset of baroclinic instability is $Ta^{\rm onset}\approx 6.33\times 10^6$ (or equivalently, $Ro_T^{\rm onset}\approx 3.1$) ; above (in terms of $Ro_T$, below) this threshold wave propagation is expected in the zonal direction (von Larcher and Egbers, 2005).

In the classic parametrization of homogeneous (i.e. thermally driven) baroclinic annulus experiments (Vincze \etal, 2014; Vincze \etal 2015) the vertical scale $h$ that appears in the definitions of non-dimensional numbers (\ref{Ta}) and (\ref{Ro}) corresponds to the total water level in the tank, since the sideways convective unicellular flow penetrates through the whole depth $D$ of the domain ($h=D$). The presence of vertical salinity stratification however, yields a system of shallow-layer convective cells (cf. Fig. \ref{dye_demo}), whose thicknesses may be determined by Chen's formula (\ref{lambda}), and by the cut-off value $\lambda_{\rm crit}$. Therefore it is useful to introduce \emph{local} versions of the Taylor and thermal Rossby numbers by taking $h=\lambda(z)$ as vertical scale, corresponding to the local $N$. Hence, the local thermal Rossby number takes the form
\begin{equation}
Ro_T(z)=\Bigg(\frac{g\alpha \Delta T}{\Omega  N(z) d}\Bigg)^2,
\label{Ro_loc}
\end{equation}
and the local $Ta(z)$ is written as
\begin{equation}
Ta(z)=\frac{4\Omega^2 d^5 N^2(z)}{\nu^2 g \alpha \Delta T}.
\label{Ta_loc}
\end{equation}
It is reasonable to assume that such parameter combinations exist, where the global Taylor number calculated with the full water depth $D$ does not exceed $Ta^{\rm onset}$, but the local Taylor number $Ta(z)$, taken with the thickness of a given convective cell ($h= \lambda(z)< D$) is already above the threshold. The multicellular state can therefore enhance baroclinic instability within the cells, and thus open the way for wave propagation in such configurations, which (given their values of $\Delta T$ and $\Omega$) would be stable in the homogeneous case. Also, due to differences in cell thicknesses $\lambda(z)$, the `baroclinicity' (i.e. the deviation of the flow field from the wave-free, axially symmetric state) in the set-up may vary from cell to cell, thus different flow states may be observed at different depths.

\section{Experimental set-up and methods}
\label{Setup}
\begin{figure}[!h]
\begin{center}
\noindent\includegraphics[width=10cm]{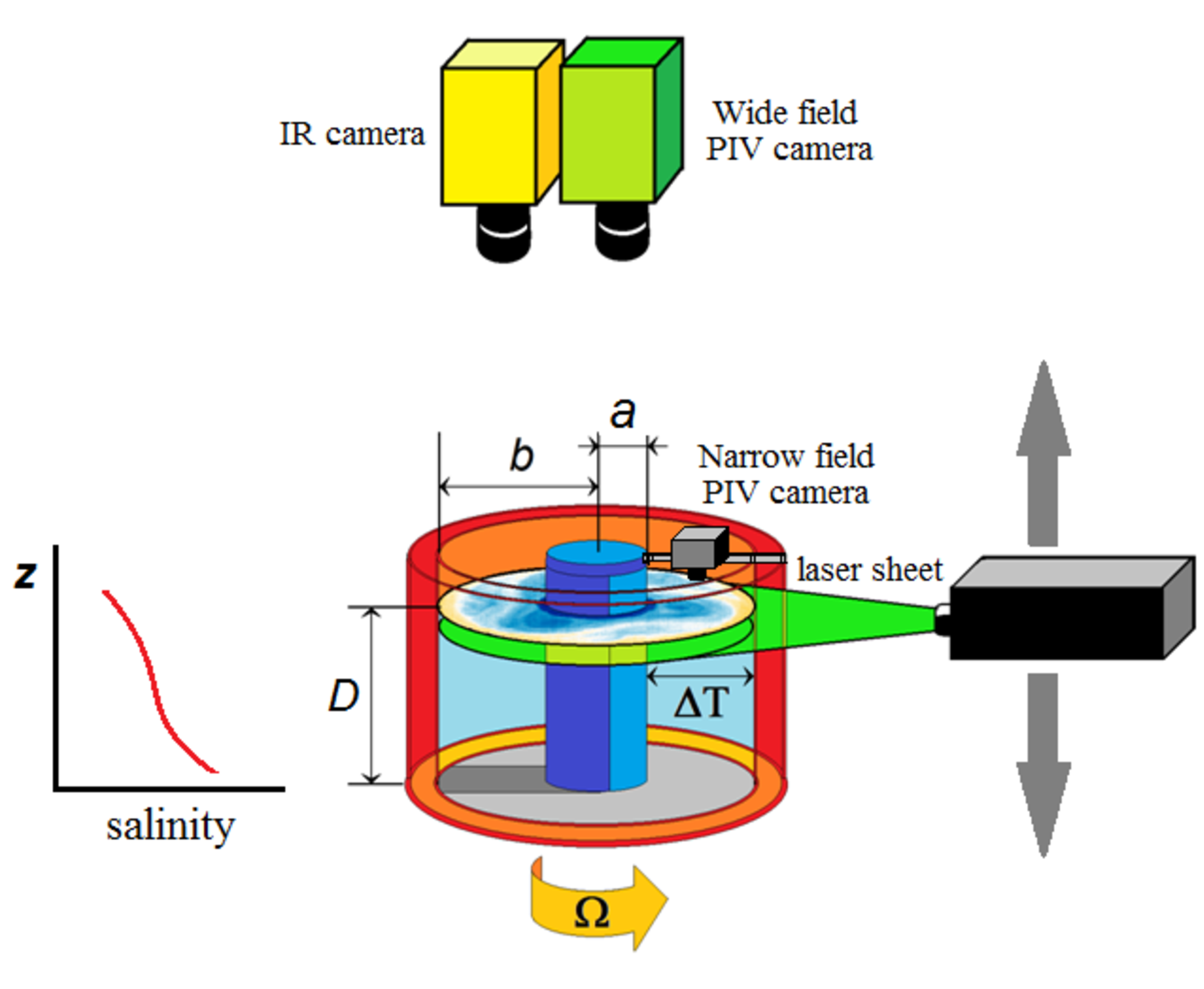}
\caption{A schematic drawing of the experimental set-up, with parameters $a=4.5$ cm, $b=12$ cm,  $D=10-13$ cm, $\Omega=1.7-2.7$ rpm, $\Delta T = 6$ K. The direction of the tank's rotation is also indicated.}
\label{setup}
\end{center}
\end{figure}

The experiments of the present study were conducted in the laboratory set-up shown in Fig. \ref{setup}. The working fluid consisted of sodium-chloride/de-ionized water solution, with Lewis number  $Le \equiv \kappa_T / \kappa_S \approx 101$, where $\kappa_T$ and $\kappa_S$ are the diffusion coefficients for heat and salt, respectively.
The radii of the inner (cooling) and outer (heating) cylinders were $a=4.5$ cm and $b=12$ cm, yielding a $d\equiv b-a=7.5$ cm wide annular cavity that was filled up to heights ranging from $D=10$ cm to $D=13$ cm in the different runs. The lateral temperature difference $\Delta T$ was set between 6 and 10 K in all cases.  A continuously stratified salinity profile was prepared in the cavity before each measurement with the standard two-bucket technique (Oster and Yamamoto, 1963) two vessels containing saline and fresh water are connected with a pipe so that the freshwater inflow to the saline bucket yields a mixture of ever decreasing salinity in time. By raising or lowering one of the buckets during this filling-up procedure, we were able to adjust the water fluxes into the mixing compartment and, thus, create  arbitrarily non-uniform stable salinity profiles.

Our experiments investigated variations of the flow field with height for different types of initial salinity stratification.
Thus, the fluid was seeded with $100 \mu$m - diameter tracers for particle image velocimetry (PIV). After filling up the tank, the initial salinity profiles were measured using a hand-held conductivity sensor. Then the differential heating was switched on.

\begin{figure}[!h]
\begin{center}
\noindent\includegraphics[width=8cm]{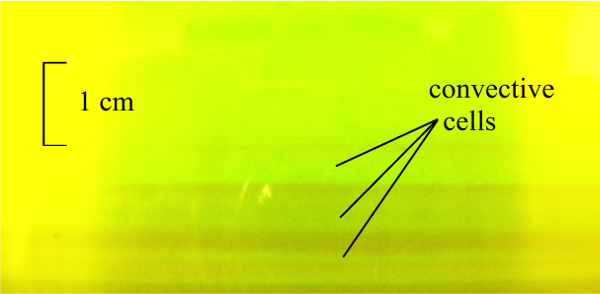}
\caption{A snapshot of `classic' double diffusive cells at the bottom of the experimental tank during a non-rotating control experiment, visualized with rhodamine. Note, that the cells do not invade the regions farther above the boundary, where $N$ is larger.}
\label{no_rot_cells}
\end{center}
\end{figure}

Even before starting rotation, the onset of double diffusive convection could be observed visually. Generally, most of the $N$-values corresponding to the prepared vertical profiles yielded smaller $Ra_{\rm Chen}$'s than the aforementioned critical threshold of convection in a non-rotating system, $Ra_{\rm Chen}^{\rm crit}\approx 15000$. Therefore, the convective flow in the bulk was mostly inhibited. However, the no-flux conditions for salinity at the bottom and the top of the water body translate to $N \approx 0$ conditions at these regions in terms of buoyancy frequency. Thus in thin layers at the vicinity of these boundaries where $N < N_{\rm crit}$ holds, localized convective cells could still partially invade the water column, as seen in Fig. \ref{no_rot_cells}, visualized with fluorescein dye.

Rotation rate $\Omega$ was increased gradually before reaching its target value (in increments of $\Delta \Omega \approx 0.2$ rpm in every 5 minutes). At this point the system was left undisturbed for 30-40 minutes, thus the early transients of the rotating double-diffusive cell formation were not observed.
Then, PIV measurements were conducted using horizontally illuminated laser sheets at different water depths. Two slightly different arrangements were used for the two sets of experiments. For the first set, a camera together with a laser were fixed on an arm from the sidewall of the container (`narrow field' observations). Close-ups of the velocity fields could be acquired in this manner. In the second set of experiments, complete `wide field' views of the velocity fields in the full annulus could be observed from a co-rotating camera platform situated above the tank (as depicted in Fig. \ref{setup}). Alongside these PIV cameras, for both sets of experiments an infrared thermographic camera was also mounted on the platform to enable the simultaneous observation of the temperature field at the free water surface. At the end of the PIV measurements, rotation rate $\Omega$ was lowered gradually (in the same manner as in the spin-up phase), and after the flow reached a full stop, a second salinity profile was also recorded.

\section{Results}
\label{Results}

\subsection{General description of the unstable flow}

\begin{figure}[!h]
\begin{center}
\noindent\includegraphics[width=10cm]{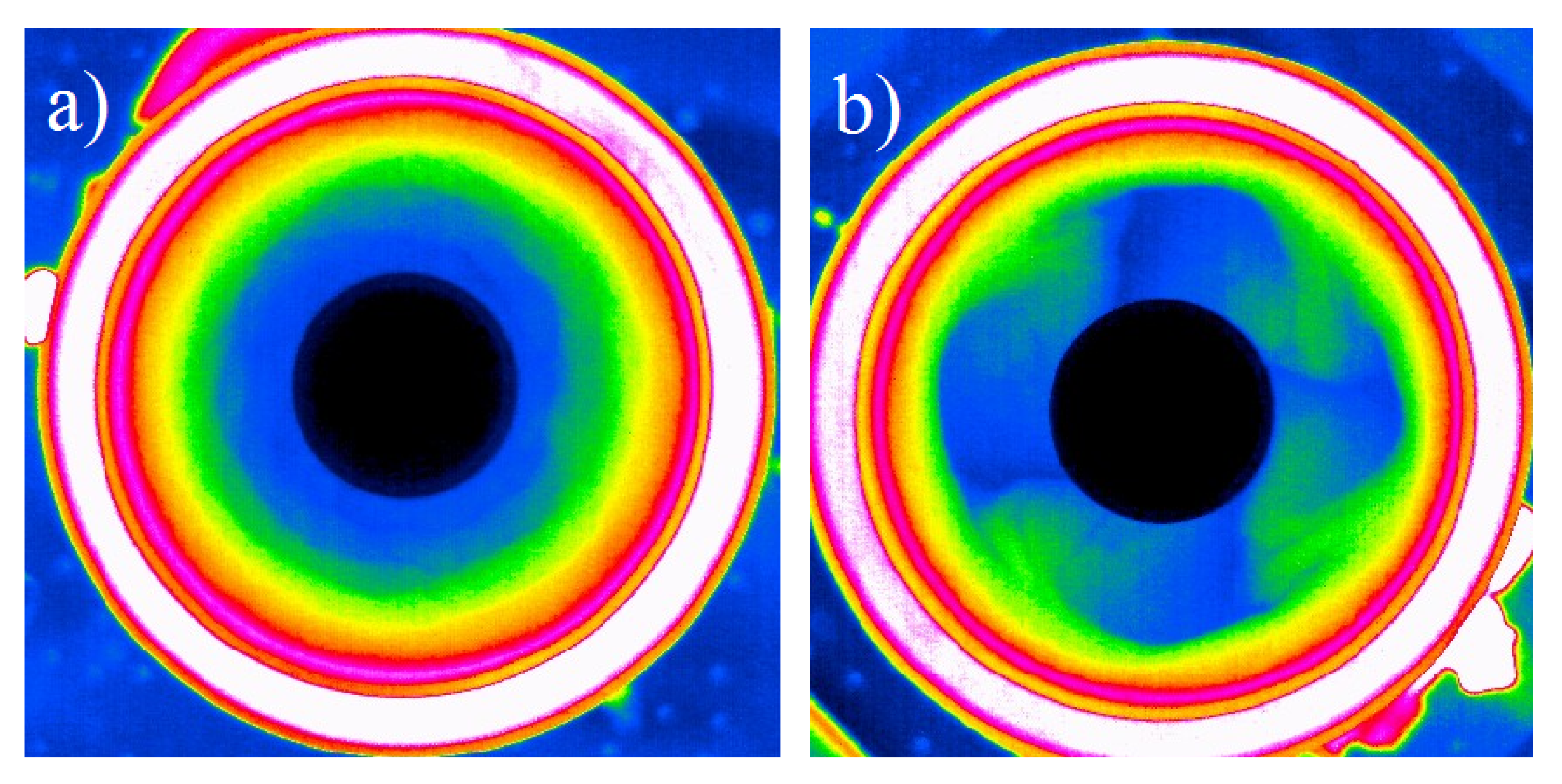}
\caption{Thermographic images of the surface temperature patterns of a homogeneous (a) and a salinity stratified (b) experimental run (in the latter case: $\langle N \rangle = 2.5$ rad/s) with the same `global' parameters: $\Delta T = 6$ K, $D = 12$ cm, and $\Omega = 2.5$ rpm.}
\label{infrared}
\end{center}
\end{figure}

The observation of surface temperature patterns in the tank using infrared camera revealed that -- contrarily to the initial na\"ive expectations of the authors, but in full agreement with our hypothesis of baroclinic destabilization via the scale selection of $\lambda$, discussed subsection \ref{DDconv} -- in experiments with salinity stratification the layer at the water surface becomes baroclinic unstable at much lower rotation rates $\Omega$ than in the homogeneous system.
This phenomenon is demonstrated in Fig. \ref{infrared}. In panel a) the surface temperature field of a homogeneous control experiment is shown, where the vertical scale is set by the total water depth $D = 12$ cm. The standard temperature difference $\Delta T = 6$ K and the applied rotation rate $\Omega = 2.5$ rpm yields $Ta = 5.3\times 10^6$, that is below the aforementioned $Ta^{\rm onset}\approx 6.33\times 10^6$, or equivalently, the corresponding $Ro_T$ is above $Ro_T^{\rm onset}\approx 3.1$, as determined by von Larcher and Egbers (2005). Consistently, the temperature distribution is axisymmetric indeed, implying that the fluid stays in the stable regime.

Panel b) shows the temperature pattern emerging in the stratified case for the same setting of parameters $D$, $\Delta T$ and $\Omega$, resulting in a fourfold symmetric baroclinic wave, generally referred to as $m = 4$ in the literature. von Larcher and Egbers (2005) found -- using the same experimental tank and $\Delta T$ as here, but with homogeneous working fluid -- that the lower limit of the $m = 4$ regime in terms of Taylor number is around $Ta \approx 10^8$ (or somewhat below, at $Ta \approx 7\times 10^7$ if `spin-down' initial conditions are applied). From the formula (\ref{Ta}) one can therefore calculate that the maximum characteristic vertical scale of the convection has to be approximately $h \approx 1$ cm to explain the observed wave number. Taking this $\lambda = h$ value as the Chen scale of (\ref{lambda}) yields a local buoyancy frequency $N\approx 1$ rad/s. This is consistent with the measured salinity profile, although -- due to the finite size of used the hand-held conductivity sensor -- there are no measured data points from the uppermost layer of the initial profile. Yet, based on the fact that 2 cm below the surface $N = 2.5$ rad/s was measured, and on the aforementioned no-flux condition for salinity prescribing $N=0$ at the top, it is reasonable to interpolate $N\approx 1$ rad/s for the considered region. Moreover, $\lambda \approx 1$ cm also matches the observed thickness of the spontaneously forming convection cells at the boundaries, cf. Fig. \ref{no_rot_cells}.

In order to obtain a more quantitative relationship between $N$ and the `baroclinicity' of the flow, one needs to investigate the bulk of the working fluid, where the buoyancy frequency profiles can be properly determined. This domain, however, is not accessible for infrared thermography (since the penetration depth of the applied infrared wavelength range in water is of the order of a millimeter). Thus, the buoyancy frequencies directly could only be compared with PIV data.

\subsection{Two separated convective layers}
\label{simple}
\begin{figure}[!h]
\begin{center}
\noindent\includegraphics[width=8cm]{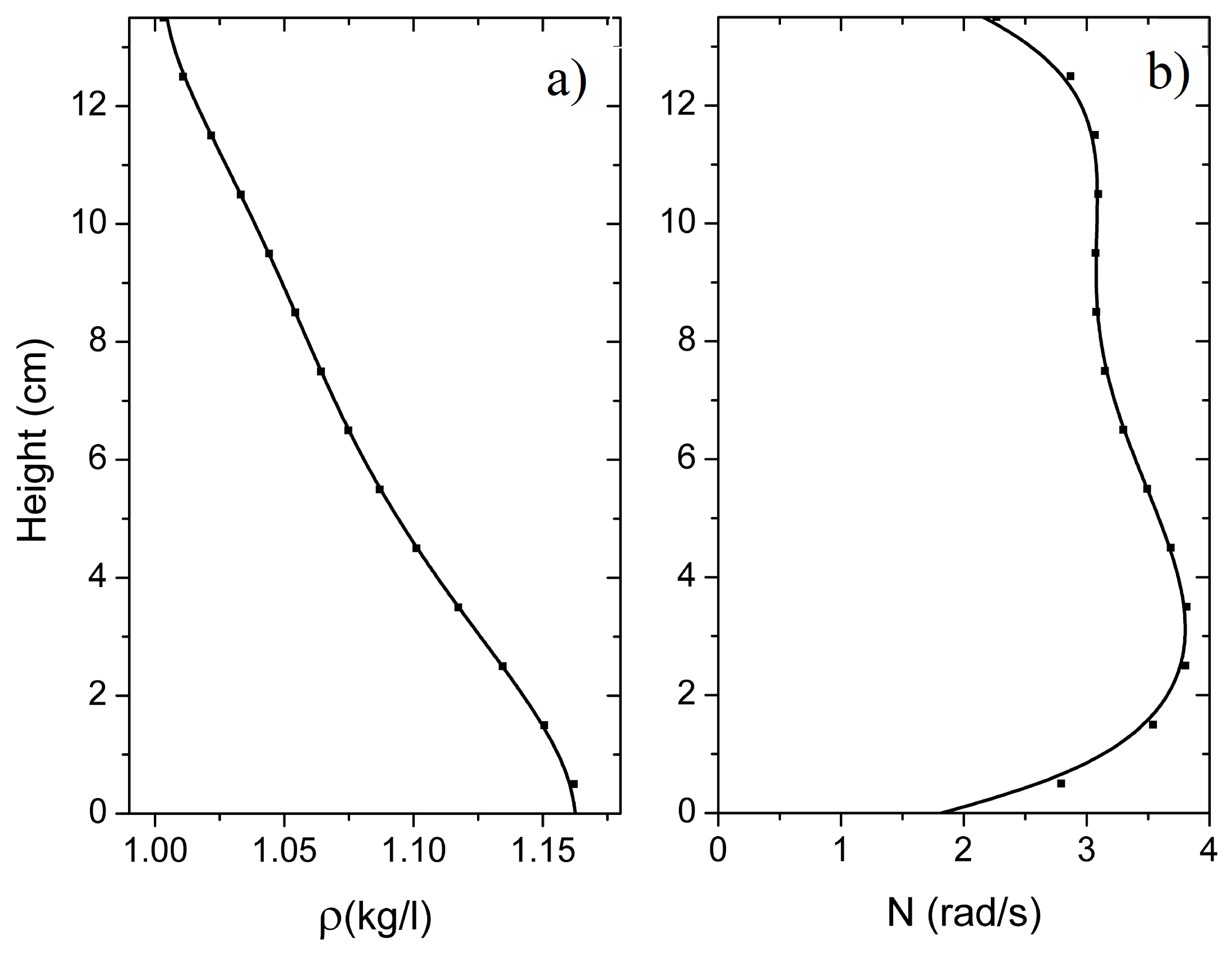}
\caption{ Saline water density (a) and the corresponding vertical buoyancy frequency profile (b) before starting the series of experiments. Visibly, except
 for the top and bottom, where the no-flux conditio yields weakly stratified layers, $\langle N \rangle$ is fairly constant around a value of $\langle N \rangle =3.3$ rad/s}
\label{Nprofileexp4}
\end{center}
\end{figure}

In this section we will present experiments which we started from  particularly simple density profiles as illustrated in figure \ref{Nprofileexp4}a.  The buoyancy frequency $N$ is plotted  versus the vertical coordinate in figure \ref{Nprofileexp4}b. We performed different experiments at the following values of global parameters: $\Delta T=10$ K and $\Omega=2$ rpm (denoted as Exp. 4), $\Delta T=9.6$ K and $\Omega=3$ rpm (Exp. 5), $\Delta T=9$ K and $\Omega=4$ rpm (Exp. 6), $\Delta T=9$ K and $\Omega=5$ rpm (Exp. 7). In the most detail the first of these runs (i.e. `Exp. 4') will be discussed.
One can see in fig. \ref{Nprofileexp4} that the vertical density profile is almost linear. At the upper and lower boundaries the no-flux boundary condition imposes constant concentration and therefore constant density. Thus, close to these boundaries $N$ approaches zero.

\begin{figure}[!h]
\begin{center}
\noindent\includegraphics[width=10cm]{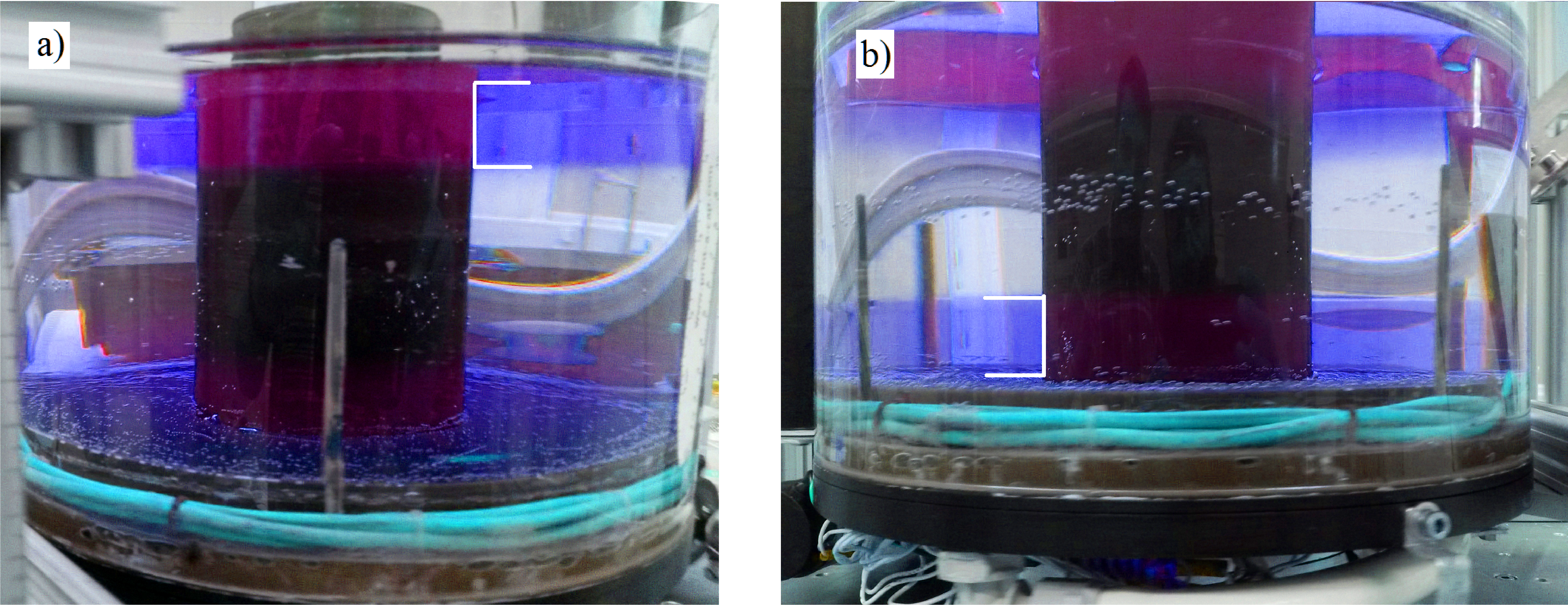}
\caption{Visualization of the convective cell confined in the upper (a) and lower (b) layers of the fluid. In the middle region no convection occours.}
\label{visuconv}
\end{center}
\end{figure}

Contrarily to the non-stratified case, as explained in section \ref{DDconv}, the imposed horizontal temperature gradient 
has to reach a certain threshold before sideways convection sets in. Figure \ref{visuconv} shows the top and bottom convective layers, where the fluid has been colored by the gentle injection of dye at the water surface. As visible and as expected, convection is confined in layers
of water where stratification is small. The thickness of this convective cell progresses slowly in time from around $2$ cm at threshold.

To obtain velocity vectors of the flow, the PIV method is suitable. As mentioned in section \ref{Setup}, in this series of experiments the narrow field camera (GoPro, 1080p, 30fps) and the co-rotating laser sheet were utilized to acquire video information. For calculating the velocities we used successive frames taken at equally distributed time intervals with increment of $0.5$ s and we processed them with the public license MatPIV toolbox. The advantages of using a co-rotating measurement system are that the camera can be positioned near the free surface and thus the resolution will increase, and we don't need to correct the calculated velocities by subtracting the tank's rotation speed.
As soon as rotation is started, a zonal flow is created by the action of the Coriolis force on the convective currents. 
 
\begin{figure}[!h]
\begin{center}
\noindent\includegraphics[width=8cm]{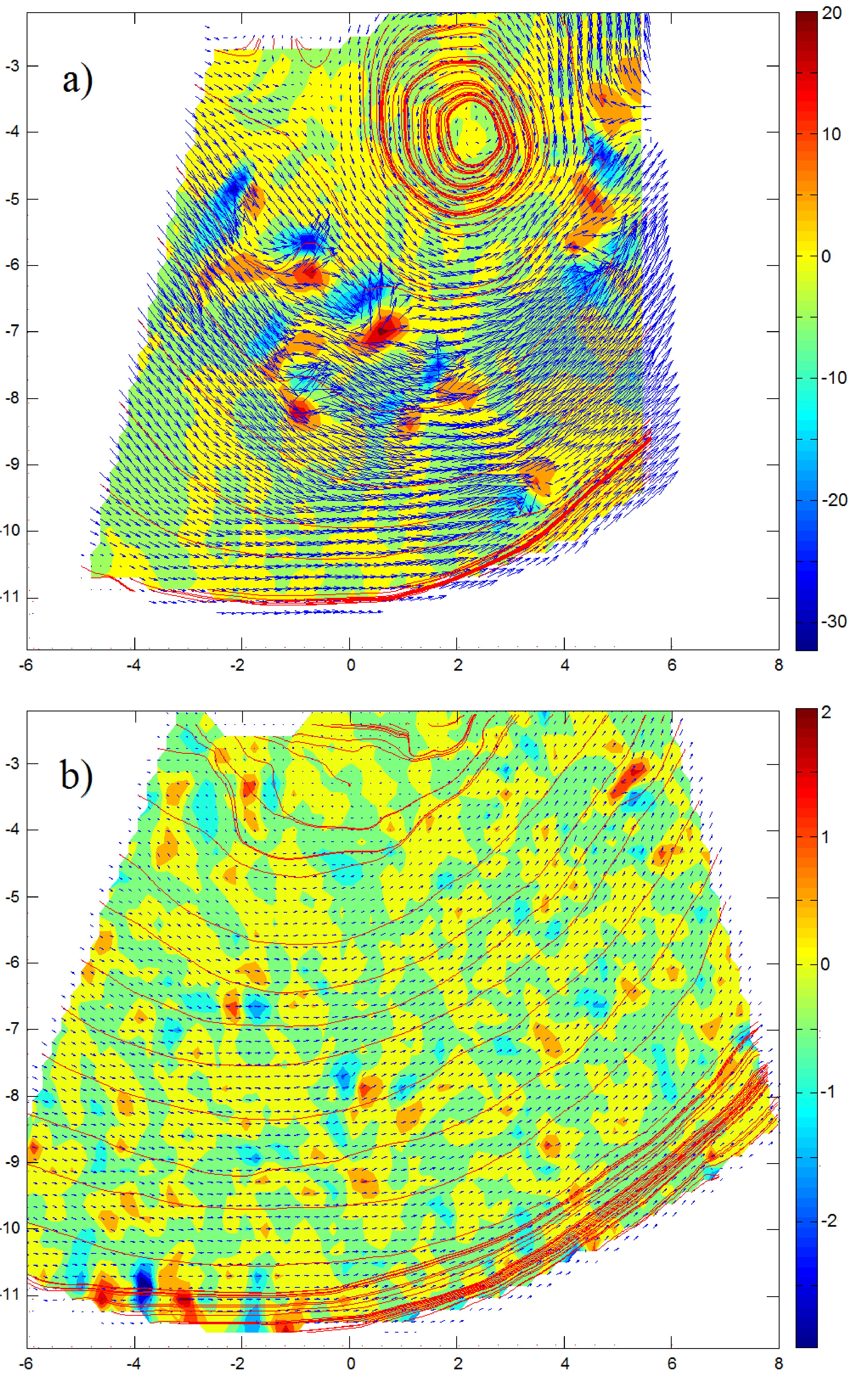}
\caption{PIV measurements of the flow at two different heights (near the surface at $z=12.5$ cm and in the middle region at $z=9$ cm) showing the azimuthal flow with or without the baroclinic instability. The small arrows represents the velocity vectors in the horizontal plane, the colored maps are the horizontal divergence fields. The stream lines are plotted in red. Close to the surface the velocities are an order of magnitude higher ($v_{max} \approx 8$ mm/s) than in the center ($v_{max}\approx 1.1$ mm/s).}
\label{PIVmap}
\end{center}
\end{figure}

PIV maps have been acquired for $9$ height levels. Depending on the rotation rate, and also on the intensity of the convective velocities, these zonal flows can be either stable or unstable in terms of baroclinic instability. Fig. \ref{PIVmap} illustrates two horizontal close-ups of the velocity fields.
The first one (panel a) is taken near the surface, inside the layer where convection is allowed to set in. Here a baroclinic wave can be observed. At the convective region close to the bottom of the tank a convective cell appears too, but in contrast with the one near the surface it is baroclinic stable, possibly because the zonal flow is weakened by the bottom boundary layer.

The next question is then to explore the possible generation of Inertial-Gravity Waves (IGWs) by the instability as it can be the case for instance in unbalanced dynamics of atmospheres; see for instance O'Sullivan and Dunkerton (1995) or Plougonven and Snyder (2007). For the first time Jacoby \etal (2010) looked for IGWs in a classical continuously stratified, differentially heated rotating annulus. They found signatures for high-frequency waves but surprisingly the waves were related to boundary layer instabilities and not to the fronts of baroclinic waves. Recently, Borchert \etal (2014) and Randriamampianina and del Arco (2015) numerically investigated the occurrence of IGWs by the baroclinic instability. The former claimed to have found IGWs generated spontaneously in accordance with the waves detected in a two-layer baroclinic annulus by Lovegrove \etal (2000) and Williams \etal (2003). The latter showed that in the continuously stratified case with small Prandtl number, IGWs are generated by Kelvin-Helmholtz instability and not via boundary layer instability. Although we cannot prove yet that the present instability can generically generate IGWs, the velocity field of Fig. \ref{PIVmap}a exhibits a wave train whose characteristics are compatible with IGWs.

The second close up (panel b) corresponds to the calm region in the center of the tank, where convection is blocked due to the effect of the strong initial stratification. Driven by the strong flow in the upper and lower convective layers there is still some azimuthal symmetric flow, but the velocities are an order of magnitude lower than the ones from the upper and lower convective layers.

\begin{figure}[!h]
\begin{center}
\noindent\includegraphics[width=10cm]{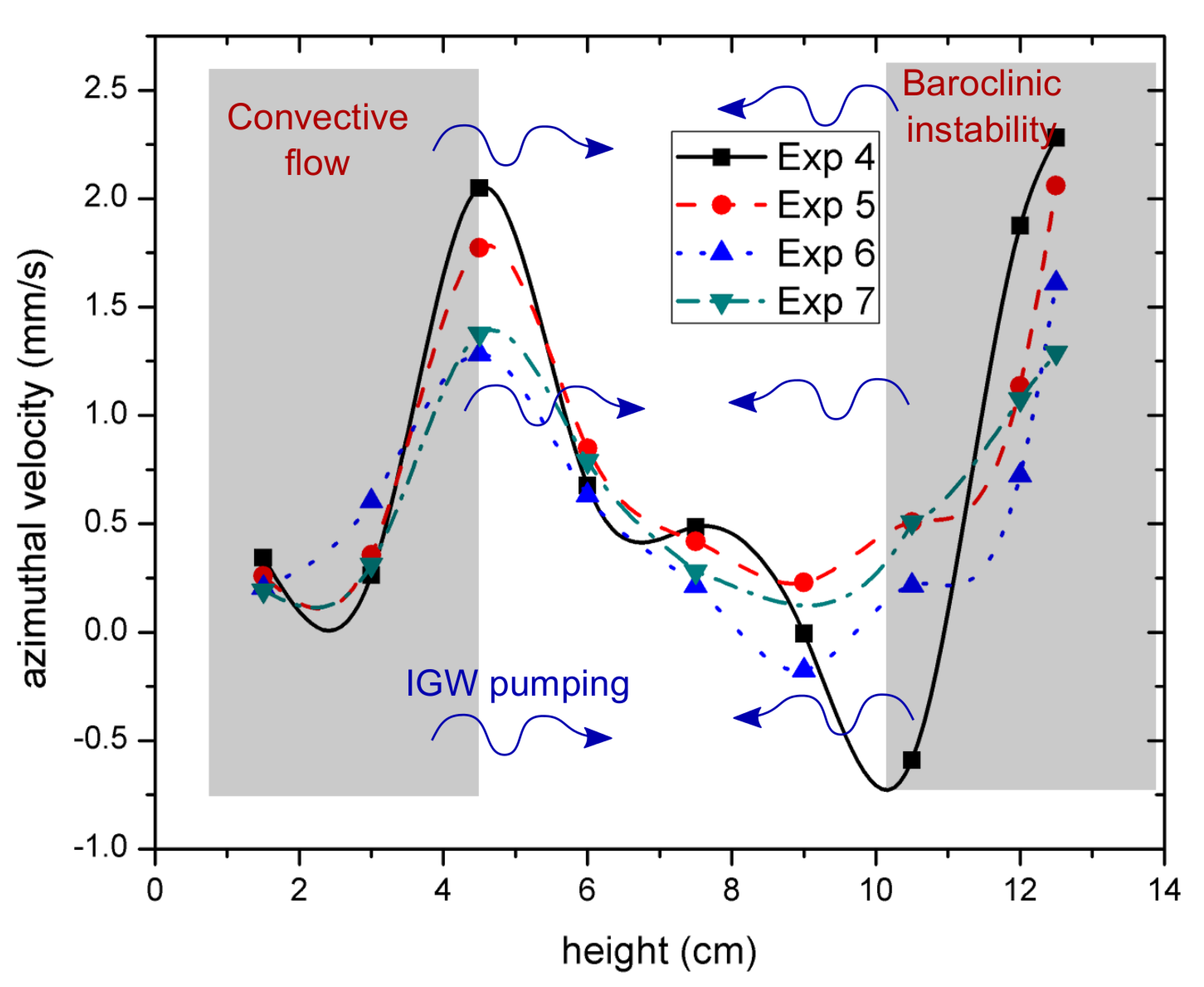}
\caption{Vertical azimuthal velocity profile as calculated from four experiments corresponding to almost the same $\Delta T$ of about $10$ K and different values of $\Omega$ showing three distinct zones: two unstable convective cells at bottom ant top 
of the full layer and a motionless layer that separate them in the middle plane.}
\label{PIVmean}
\end{center}
\end{figure}

This flow structure demonstrates the features that was illustrated
by our numerical observation presented in section \ref{DDconv}. Fig. \ref{PIVmean} presents the vertical azimuthal velocity
 profile as measured in the experimental runs denoted Exp.4 to Exp.7, as mentioned above. For each point we calculated the mean velocity over the whole investigated plane, which corresponds to approximatively one sixth of the cylindrical gap in the azimuthal direction. Since the presence of baroclinic instability breaks the axial symmetry of the flow, temporal averaging was performed upon a sufficient timespan to permit that the flow from different segments can contribute to the final azimuthal mean (between one and $2.5$ minutes). Although the vertical resolution is rather poor ($9$ data points in the
 full water layer) one can easily recognize the opposite signs of the zonal flows at the borders of the convective cells. In the small region between them only weak flow was measured.  It is to be emphasized, that despite the differences in rotation rate $\Omega$ and temperature difference $\Delta T$, all four experiments yielded very similar results. It is also worth mentioning here that this motionless layer can possibly sustain
inertia-gravity waves (IGWs) that would be launched from the top or the bottom unstable zones.

\subsection{Multilayered convection}
\label{complex}

\begin{figure*}[ht]
\begin{center}
\noindent\includegraphics[width=11cm]{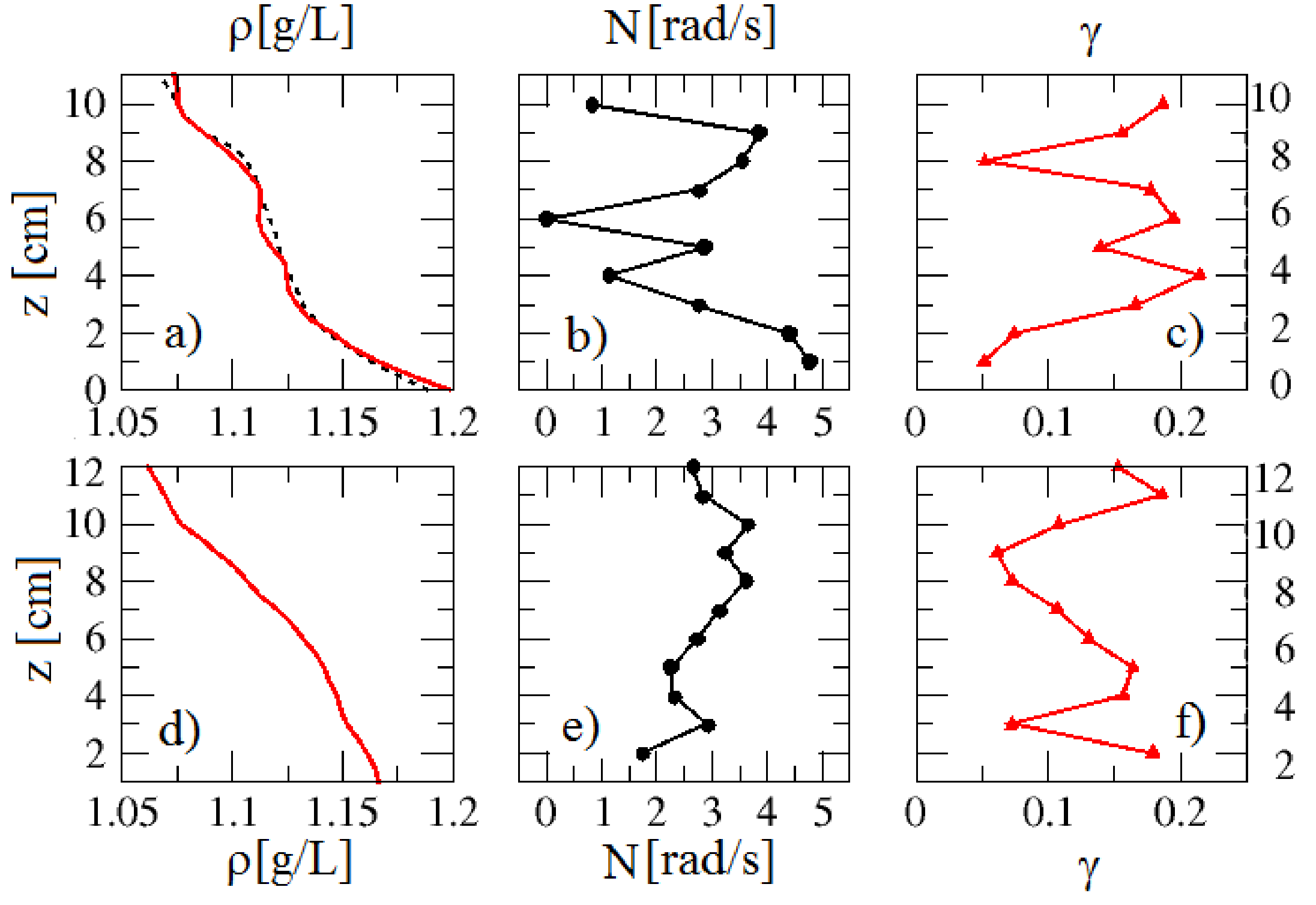}
\caption{Vertical profiles of density (a and d), buoyancy frequency (b and e) and anisotropy parameter $\gamma$, defined in (\ref{gamma}), for two experimets. The applied rotation rates were $\Omega = 2.24$ rpm for the upper, and $\Omega = 1.7$ rpm for the bottom row ($\Delta T = 6$ K).}
\label{profiles}
\end{center}
\end{figure*}

\begin{figure}[h]
\begin{center}
\noindent\includegraphics[width=7cm]{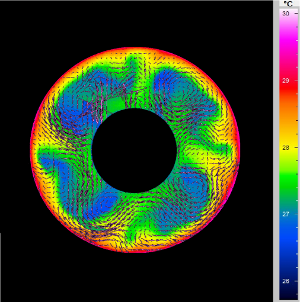}
\caption{Surface temperature (color) and simultaneous horizontal velocity fields (arrows) from the uppermost 1 cm layer of one of the multilayered experimental runs ($\Delta T = 6$ K, $\Omega = 2.7$ rpm).}
\label{piv_field}
\end{center}
\end{figure}

In this section we present our findings from the second set of experiments, where more complex initial salinity profiles were studied, two of which are presented in panels a) and d) of Fig. \ref{profiles}. In panel a) both the initial (dashed line) and final (solid line) profiles are shown. Similarly to the simulated final profile in Fig. \ref{sim_profiles}, the presence of double-diffusive staircase is apparent. Panels b) and e) show the values of buoyancy frequency $N$, as calculated from polynomial fits of the density profiles, and evaluated at the height levels where the `wide field' PIV measurements were conducted.
As an example of the PIV fields evaluated in this series of experiments Fig. \ref{piv_field} shows the acquired velocity vectors of a fully developed fourfold symmetric wave pattern obtained from ca. 1 cm below the water surface, superimposed onto the simultaneously obtained surface temperature field.  

\begin{figure*}[ht]
\begin{center}
\noindent\includegraphics[width=10cm]{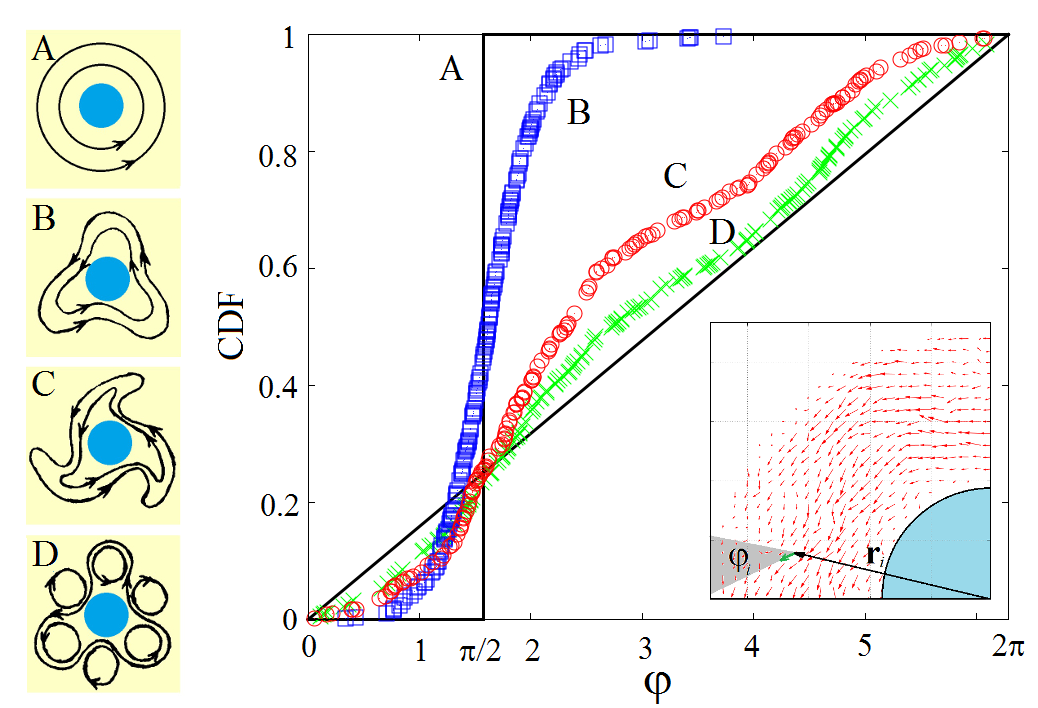}
\caption{The development of baroclinic instability towards larger values of $Ta$ (smaller values of $Ro_T$). Sketches A-D (left) are cartoons of the streamlines in the flows to which the CDFs of the main panel belong. The right hand side shows the corresponding CDFs of velocity angle $\varphi$ as obtained from actual PIV measurements. The inset demonstrates the geometric interpretation of angles $\varphi_i$.}
\label{cdf_fig}
\end{center}
\end{figure*}

At each of these levels, the horizontal velocity field was calculated on a lattice of $49 \times 66$ grid points, yielding a spatial resolution of around $0.6$ cm. In order to obtain a proper measure of `baroclinicity' in such a slice of the flow, firstly, the angle $\varphi_i \in [0;2\pi)$ between each velocity vector and the respective position vector $\vec{r}_i$ -- directed from the center of the tank, i.e. the axis of rotation -- was computed (see the explanatory sketch in the inset at the bottom right corner of Fig. \ref{cdf_fig}). Our approach is based on the distribution of these $\varphi_i$-values. In the absence of baroclinic instability, the temperature pattern is axially symmetric and the flow is zonal (see Fig. \ref{infrared}a or state A in Fig. \ref{cdf_fig}), thus the velocity field is highly {\it anisotropic} in the sense that the vast majority of the velocity vectors are perpendicular to their position vectors ($\varphi_i\approx \pi/2, \,\, \forall i$). At larger values of $Ta$ a meandering jet appears (state B in Fig. \ref{cdf_fig}): the values of $\varphi_i$ scatter in a broader range around $\pi/2$. In a fully developed baroclinic unstable case (states C and D) the flow field is dominated by vortices, making the $\varphi_i$ density distribution close to uniform within the interval $[0;2\pi)$, i.e., in a certain sense, {\it isotropic}.

The right panel shows the cumulative density functions (CDF) of the $\varphi_i$ values of four measured vector fields, whose streamlines (states B through D) were sketched, as discussed above. Each count of the CDFs was calculated with a certain weight factor, assigned to each velocity vector proportionally to their relative magnitudes compared to the mean $\langle u \rangle$ (after the careful removal of outliers), in order to compensate for the inevitable uncertainties in the $\varphi_i$ values when measuring small vectors. In a perfectly isotropic case, the CDF curve would follow the ${\rm CDF}(\varphi_i)=\varphi_i/(2\pi)$ linear function. The more anisotropic the velocity field, the larger the deviation of the ${\rm CDF}(\varphi_i)$ graph from this straight line.
We quantify the anisotropy of the flow by defining a parameter $\gamma$ as
\begin{equation}
\gamma=\frac{1}{n}\sum_i\left|{\rm CDF}(\varphi_i)-\frac{\varphi_i}{2\pi}\right|,
\label{gamma}
\end{equation}
where $n$ is the total number of the obtained `useful' velocity vectors of the PIV data, which lie within the measurement domain (the annular cavity covers only a fraction of the rectangular image) and remain after the removal of clear outliers. This process typically left us with $n\approx 850$ values of $\varphi_i$ for a given horizontal PIV slice.
$\gamma$ serves as the desired measure (order parameter) of baroclinic instability: generally, the smaller $\gamma$ is, the more baroclinic unstable the system appears to be in the considered horizontal section. A perfectly axisymmetric flow, with all the velocity vectors aligned perpendicularly to their position vectors would yield $\gamma =0.75$ (independently of whether the flow is clockwise or counterclockwise), whereas in a fully developed geostrophic turbulent state, where all directions $\varphi$ are equally probable $\gamma \approx 0$ would hold.

The vertical profiles of anisotropy parameter $\gamma(z)$ obtained for the two measurements of Fig. \ref{profiles} are shown there in panels c) and f). We emphasize that the determination of these $\gamma$-values is completely independent from that of the density profiles, the former being obtained directly from the PIV data, and the latter via a hand-held conductivity probe.
Yet, there is an apparent anticorrelation between $\gamma(z)$ and $N(z)$ in both shown cases, as seen when comparing panels b) to c), and e) to f) in Fig. \ref{profiles}.

\begin{figure*}[ht]
\begin{center}
\noindent\includegraphics[width=15cm]{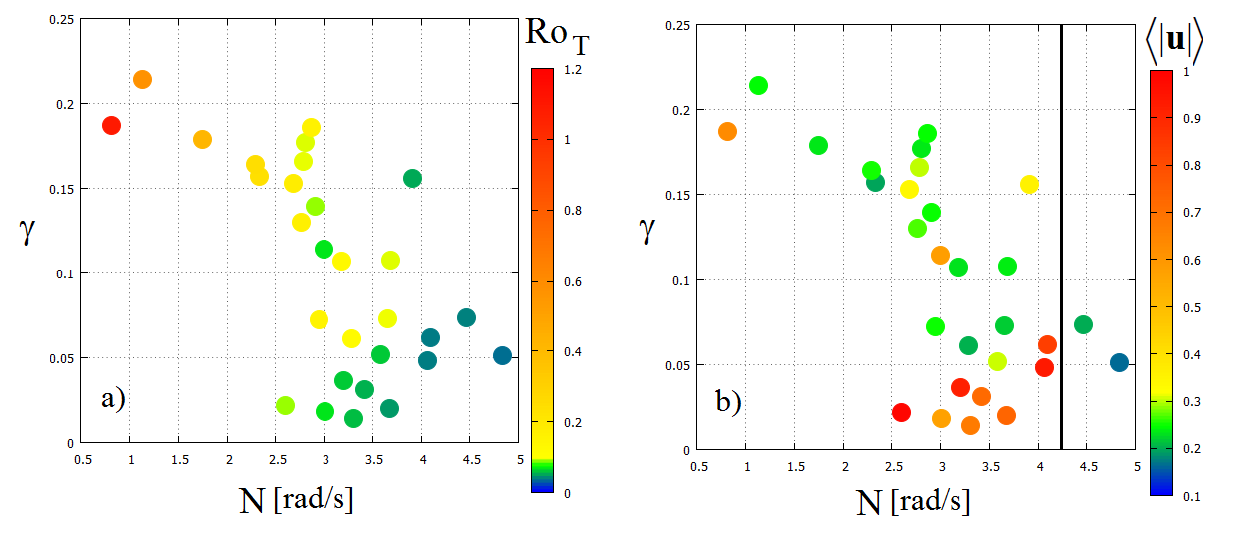}
\caption{Scatter plots of local anisotropy parameter $\gamma$ vs. buoyancy frequency $N$, obtained from four experimental runs with different density profiles and rotation rates. The color coding is based on the local thermal Rossby number $Ro_T$ (a) and the normalized average horizontal velocity $\langle|\vec{u}|\rangle$ (b). In panel b) the vertical solid line indicates  $N_{\rm crit}\approx 4.2$.}
\label{alldata}
\end{center}
\end{figure*}

The acquired $\gamma(N)$ data points from four experiments with different vertical density profiles and slightly different rotation rates (ranging from $\Omega=1.7$ rpm to $2.7$ rpm) are combined in Fig. \ref{alldata}a, with colors accounting for the corresponding values of local thermal Rossby number $Ro_T(z)$, as in (\ref{Ro_loc}): here each point is calculated from a complete horizontal PIV velocity slice. The overall trends are apparent: firstly, it is reassuring that anisotropy parameter $\gamma$ seems to behave consistently with $Ro_T$. Generally, the smaller $Ro_T$ is, the closer the system gets to geostrophic turbulence, yielding smaller $\gamma$'s, as expected. Secondly, $\gamma$ markedly anticorrelates with $N$ in itself too, despite of the differences in the rotation rates, underlining the robustness of the $N$-dependence: the steeper the stratification, the smaller the vertical convection scale becomes, supporting the development of shallow layer geostrophic turbulence, i.e. smaller $\gamma$.

Fig. \ref{alldata}b shows the same data set, now colored with respect to the corresponding relative average absolute horizontal velocities $\langle|\vec{u}|\rangle$, based on the PIV fields. These values have been normalized so that $\langle|\vec{u}|\rangle=1$ corresponds to the largest observed average velocity.
We found that two of the three weakest observed flow layers (the two blue and green data points on the right of the plot) appear at the largest buoyancy frequencies ($N>4.2$ rad/s), which is in sharp contrast with the large velocities measured at only slightly smaller $N$'s.
This may well imply that here the blocking threshold is approached, and above $N_{\rm crit}\approx 4.2$ rad/s (marked by a solid vertical line) the convection is inhibited; thus the corresponding parameters $\gamma$ do not have actual physical meaning here, being results of the angle statistics of noise-dominated PIV fields. The consistency of this finding with the theoretical background is addressed in the next section.

\section{Discussion}
\label{Discussion}

Chen \etal (1971) and Kerr (1995) have analyzed the conditions under which double diffusive convection can set on in laterally heated, initially stratified set-ups. For a non-rotating case the criterion (as reported in the experimental study of Chen \etal (1971)) was found to be that the `combined' Rayleigh number, which incorporates the effects of both the salinity and the temperature gradients as defined in (\ref{Ra_Chen}), must exceed a certain threshold $Ra_{\rm Chen}^{\rm crit}\approx 15000$. From here, one can calculate the critical buoyancy frequency $N^{\rm crit}_{\rm Chen}$ above which the viscous effects inhibit the formation of convective cells in our experimental tank. This estimation yields $N^{\rm crit}_{\rm Chen}\approx 1.7$ rad/s, which appears to be consistent with the observation that in the \emph{non-rotating} control experiments, cell formation was limited to the vicinity of the top and bottom boundaries, where $N$ is small (see Fig. \ref{no_rot_cells}), and is also in fairly good agreement with the results of the first set of our \emph{rotating} experiments, descussed in Section \ref{simple}.

Based on this, we would barely see any flow in the bulk of the fluid at all, because most of our actual $N$-values scatter in the interval $N \in (1.7; 4)$ rad/s, as shown in Figs. \ref{profiles} and \ref{alldata}. Yet, in our second set of experiments, where more complex initial salinity profiles were investigated (section \ref{complex}), once rotation was turned on, an alternating zonal flow emerged at mid-depths as well (see Fig. \ref{dye_demo}), implying convection.
Therefore it is reasonable to assume that in the rotating case the system can also exhibit another type of dynamics, which we referred to as the Kerr branch. The states of this branch only exist in the presence of rotation, as seen in Fig. \ref{bifurc}.

Kerr's analysis is based on a horizontal Rayleigh number $H\equiv {g\alpha \Delta T d^3}/(\nu \kappa_T)$ and a vertical one
$R\equiv {N^2 d^4}/(\nu \kappa_T)$, as discussed in subsection I.B.
The average $N$ of our  multilayered  experiments yields $R\sim 10^8$, and their typical rotation rate $\Omega \sim 2$ rpm translates to $f \sim 2 \times10^4$ in the dimensionless units applied in the analysis (using $d^2/\kappa_T$ as time units, cf. Kerr (1995)). From here, using the data presented by Kerr (1995), we can also estimate that $H_{\rm crit}\sim 10^6$ is the corresponding critical horizontal Rayleigh number. Knowing now the ratio $H_{\rm crit}/R\sim 10^{-2}$, we can directly obtain the actual critical Chen scale as $\lambda_{\rm crit}=(H_{\rm crit}/R)d$, and finally, applying (\ref{lambda}), we arrive at the theoretical prediction of $N_{\rm crit}$ in the form of:
\begin{equation}
N_{\rm crit}^{\rm Kerr}=\sqrt{\frac{g \alpha \Delta T}{(H_{\rm crit}/R)d}}\approx 4\, {\rm rad/s},
\end{equation}
that is in perfect agreement with the observed $N_{\rm crit}\approx 4.2$ rad/s, implying that our multilayered experiments followed the `Kerr (rotating) branch' of solutions, depicted in Fig. \ref{bifurc}.

Note, however, that -- as mentioned above -- our first set of experiments, where only two separate convective layers were observed at the small-$N$ regions of the top and bottom of the tank, convection was blocked already at $N\approx 3.3$ rad/s, and remained blocked, even when higher values of $\Delta T$ and $f$ were applied. This observation implies that the dynamics in these experiments stayed on the `Chen branch' of Fig. \ref{bifurc}. The present study lacks an empirical determination of the breaking point $f'$, where -- according to Kerr's linear stability analysis -- the Chen branch is expected to vanish and the Kerr to remain the only possible solution. This transition between the two types of solutions needs to be clarified by future experiments.   

\section{Conclusion}
\label{Conclusion}

To the best of our knowledge, we have reported here the very first experimental results on the combined effect of double convective and baroclinic instabilities in a rotating stratified layer and we decided to coin this phenomenon the `barostrat instability'. Following the previous analyses of Chen \etal (1971) and Kerr (1995) we have observed the formation of patterns with different azimuthal wave numbers.  As expected, these structures are confined in layers where the initial vertical density stratification was smaller.  The thresholds for convective cell formation in terms of buoyancy frequency $N$ and the observed typical vertical length scales were found to be in agreement with the theoretical predictions. 
We hope that this study will open new routes in the study of atmospheric baroclinic instabilities: in particular let us remark that the stable stratified zone that lies under the unstable shallow layer in the experiment mimics the stratosphere above the baroclinic unstable layers of the troposphere. The next question is then to explore the possible generation of inertial-gravity waves by the barostrat instability as it can be the case for instance in the unbalanced dynamics of atmospheric fronts. 

As closing remarks, we emphasize that it was somewhat counter-intuitive for the authors to find that the presence of moderately strong vertical salinity stratification can increase isotropy, i.e. enhance the development of baroclinic instability in certain shallow layers via the formation of double-diffusive cells. This is seemingly in contrast with reasoning that stable stratification (large $N$) generally blocks baroclinic instability. Our findings imply that -- although, indeed, the unicellular full-depth overturning may be inhibited -- the steeper the stratification is, the shallower the double-diffusive cells may be, which then decreases the local value of thermal Rossby number $Ro_T$, and by doing so, pushes the system (i.e. the given sideways-convective layer) towards quasi-geostrophic turbulence, as long as the stratification does not exceed a critical value $N_{\rm crit}$ above which the flow is indeed completely blocked.

With presenting the new findings of the present pilot study (which clearly needs to be followed by further experimental and numerical investigations) the authors hope to increase awareness within the community of the fact that the velocity field, and thus, horizontal {\it mixing} can vary in such a nontrivial way with $N$. This may be of interest in ocean dynamics, where horizontal (e.g. meridional) temperature difference and vertical salinity gradients co-exist.

\subsection*{Acknowledgments}
This work was supported by the European High-performance Infrastructures in Turbulence (EuHIT) program and by OTKA Grant No.
NK100296. The authors are grateful for Christian Maa{\ss}, Yongtai Wang, and Torsten Seelig for the technical support during the measurement campaigns. The fruitful discussions with Tamas Tel and Yossi Ashkenazy are also highly acknowledged. 

\section*{References}
\begin{harvard}
\item[] Boehrer B 2012 Double-Diffusive Convection in Lakes {\it Encyclopedia of Lakes and Reservoirs} 223-224

\item[] Borchert S, Achatz U, and Fruman M D 2014 Gravity wave emission in an atmosphere-like configuration of the differentially heated rotating annulus experiment {\it J. Fluid Mech.} {\bf 758} 287-311

\item[] Chen C Briggs D and Wirtz R 1971 Stability of thermal convection in a salinity gradient due to lateral heating {\it International Journal of Heat and Mass Transfer} {\bf 14} 57-65

\item[] Fowlis W and Hide R 1965 Thermal convection in a rotating annulus of liquid:
effect of viscosity on the transition between axisymmetric and non-axisymmetric flow regimes
{\it J. Atmos. Sci.} \textbf{22} 541-558

\item[] Fultz D, Long R, Owens G, Bohan W, Kaylor R and Weil J 1959
Studies of thermal convection in a rotating cylinder with implications for atmospheric motions
{\it Meteor. Monogr. Amer. Meteor. Soc.} \textbf{4} 1-104

\item[] Jacoby T N L, Read P L, Williams P D, and Young R M B 2010, Generation of inertia-gravity waves in the rotating thermal annulus by a localized boundary layer instability {\it Geophys. Astrophys. Fluid Dyn.} {\bf 105} 161-181

\item[] K\"ampf J 2010 {\it Advanced Ocean Modelling Using Open-Source Software} (Berlin-Heidelberg: Springer verlag)

\item[] Kerr O 1995 The effect of rotation on double-diffusive convection in a laterally heated vertical slot {\it J. Fluid Mech.} {\bf 301} 345-370

\item[] Kranenborg E and Dijkstra H 1995 The structure of (linearly) stable double diffusive flow patterns in a laterally heated stratified liquid {\it Phys. Fluids} {\bf 7} 680 
  
\item[] von Larcher T and Egbers C 2005 Experiments on transitions of baroclinic waves in a differentially heated rotating annulus {\it Nonlin. Processes Geophys.} {\bf 12} 1033-1041

\item[] Lovegrove A F, Read P L, and Richards C J 2000 Generation of inertia-gravity waves in a baroclinically unstable fluid {\it  Quart. J. Roy. Meteor. Soc.} {\bf 126} 3233-3254

\item[] Nettelmann N, Fortney J, Moore K and Mankovich C 2015 An exploration of double diffusive convection in Jupiter as a result of hydrogen-helium phase separation {\it Monthly Notices of the Royal Astronomical Society} {\bf 447} 3422–3441

\item[] Oster G and Yamamoto M 1963 Density gradient techniques 
{\it Chem. Rev.} {\bf 63} 257-268

\item[] O'Sullivan D and Dunkerton T 1995 Generation of inertia-gravity waves in a simulated life cycle of baroclinic instability {\it J. Atmos. Sci.} {\bf 52 (21)} 3695

\item[] Plougonven R, Snyder C 2007 Inertia-Gravity Waves Spontaneously
Generated by Jets and Fronts. Part I: Different Baroclinic Life Cycles {\it J. Atmos. Sci.} {\bf 64} 2502–2520

\item[] Randriamampianina A and del Arco E C 2015 Inertia–gravity waves in a liquid-filled, differentially heated, rotating annulus {\it J. Fluid Mech.} {\bf 782} 144-177

\item[] Schmitt R 1994 Double diffusion in oceanography {\it Annual Review of Fluid Mechanics} {\bf 26} 255-285

\item[] Vallis G K 2006 {\it Atmospheric and Oceanic Fluid Dynamics - Fundamentals and Large-Scale Circulation}, 
(Cambridge: Cambridge University Press)

\item[] Vincze M, Harlander U, von Larcher T and Egbers C 2014 An experimental study of regime transitions in a differentially heated baroclinic annulus with flat and sloping bottom topographies {\it Nonlin. Processes Geophys.} {\bf 21} 237-250

\item[] Vincze M, Borchert S, Achatz U, von Larcher T, Baumann M, Liersch C, Remmler S, 
Beck T, Alexandrov K, Egbers C, Froehlich J, Heuveline V, Hickel S and Harlander U 2015
Benchmarking in a rotating annulus {\it Meteorol. Z.} \textbf{23} 611-635

\item[] Williams P D, Read P L, and Haine T W N 2003 Spontaneous generation and impact of inertia-gravity waves in a stratified, two-layer shear flow {\it Geophys. Res. Letters} {\bf 30} 2255.   
\end{harvard}
\end{document}